\input mn
\input psfig.tex

\def\frac#1#2{{\begingroup#1\endgroup\over#2}}

\begintopmatter
\title{Towards a Direct Detection of Warm Gas in Galactic Haloes at 
Cosmological Distances}

\author{Milan M.~\'{C}irkovi\'{c},$^{1,2}$
        J.~Bland-Hawthorn$^3$ and
	Srdjan Samurovi\'c$^{4,5}$}

\affiliation{$^1$ Dept. of Physics \& Astronomy, 
SUNY at Stony Brook, 
Stony Brook, NY 11794-3800, USA\break
 cirkovic@sbast3.ess.sunysb.edu}

\smallskip
\affiliation{$^2$ Astronomical Observatory, Volgina 7, 11000 Belgrade, SERBIA}

\smallskip

\affiliation{$^3$Anglo-Australian Observatory, 
P.O. Box 296, 
Epping, NSW 2121, AUSTRALIA\break
 jbh@aaoepp.aao.gov.au}
\smallskip

\affiliation{$^4$Public Observatory, 
Gornji Grad 16,
11000 Belgrade, SERBIA}
\smallskip

\affiliation{$^5$Dipartimento di Astronomia, Universit\`a di Trieste, 
Via Tiepolo 11, I-34131 Trieste, ITALY\break samurovi@newton.daut.univ.trieste.it}
\smallskip

\shortauthor{M. M. \'Cirkovi\'c, J. Bland-Hawthorn and S. Samurovi\'c}

\shorttitle{Warm Gas in Galactic Haloes at Cosmological Distances}

\abstract{Recent highly sensitive detections of line emission from 
extended gas in the local universe demonstrate the feasibility of 
detecting H$\alpha$ emitting galactic halos out to $z\sim1$. We 
determine the form of the surface brightness vs.~redshift dependence 
which takes into account UV background evolution. Successful
detections will have a major impact on a wide range of fields, 
in particular, the source of ionization in QSO absorption systems.}

\maketitle

\section{Introduction}
The optical line emission of extended extragalactic photoionized 
sources has been recently considered in several both theoretical and 
observational works (Hogan \& Weymann 1987; Maloney 1992; Binette 
et al.~1993; Bland-Hawthorn et al.~1994; Donahue, Aldering \& Stocke 
1995; Bland-Hawthorn et
al.  1995; Gould \& Weinberg 1996; Bechtold et al. 1997; 
Bland-Hawthorn 1997;  Bland-Hawthorn et al. 1997; \'Cirkovi\'c \& 
Samurovi\'c 1998). One of the most 
important possibilities of such observations are direct detections of 
extended gaseous structures around normal luminous galaxies at various epochs.

Theoretical models of both galactic halo structure (Bregman 1981; Kovalenko,
 Shchekinov \& 
Suchkov 1989; Norman \& Ikeuchi 1989; Wolfire et al. 1995) and QSO absorption 
line systems (Mo  1994; Mo \& Miralda-Escud\'e 1996; Chiba \& Nath 1997) 
predict vast quantities of photoionized gas at large galactocentric distances 
(from several kpc to $\sim 10^2$ kpc). Low-redshift observations of 
 ${\rm Ly}\alpha$ 
absorbing systems reveal tenuous gas extending to $\sim 300$ kpc 
(Chen et al. 1998, Lanzetta et al. 1995). The empirical evidence for 
the metal-line absorbers residing in $\sim 50$ kpc haloes is very strong as well 
(Steidel 1993; Steidel, Dickinson \& Persson 1994). In view of recent claimed 
detections of extraplanar gas in recombination H$\alpha$ emission at large 
galactocentric distances in the local universe (Donahue et al. 1995),
the possibility of such a situation being typical for galaxies at all epochs 
must be examined. 

Hierarchical structure formation models also emphasize such a picture (Mo \&
Miralda-Escud\'e 1996). Detailed N-body simulations (e.g. Navarro \& White
1994), as well as the gasdynamical approach of Nulsen and Fabian (1997), show that
during the process of galaxy formation a halo of hot gas will inevitably form,
and subsequently cool until the cooling time becomes similar to the age of the
system. A natural consequence of these scenarios is that warm photoionized
gas in haloes will be bound to galaxies at all epochs. 

More recently, the focus of the discussion of dark matter in galaxy haloes has
returned to dark matter in the form of baryons. 
Big Bang nucleosynthesis requires much more baryons than observed in the
stars, interstellar and intracluster medium (Carr 1994; Fukugita, Hogan \& Peebles
1998). We investigate the possibility that at least a part of the 
baryonic dark matter is in the form of gas---presumably the same, or tightly 
related, gas which produces QSO absorption lines at low redshift. In the best
available baryonic census of Fukugita et al. (1998), warm ionized
gas around field galaxies is, significantly enough, the largest and {\it
simultaneously the most uncertain} entry in their low-redshift list. 

Deep optical searches, including HDF, severely limit the mass-to-light ratio of 
the dark matter in halo of our Galaxy and the Local Group (Richstone et al. 
1992; Flynn, Gould \& Bahcall 1996).  The most recent summary of
the MACHO project indicates that as much as half of the dark matter in
the Galaxy out to the LMC is made up of solar mass objects. The source
of the  missing mass is controversial, e.g. a halo population of white dwarfs 
(Adams \& Laughlin 1996; Kawaler 1996; Chabrier \& Mera 1997),
solar mass black holes (Moore 1993), and so on. One approach to
ruling out various models is monitoring halo evolution in galaxies from deep 
broadband images (e.g. Charlot \& Silk 1995).  For the want of suitable limits
or detections, these studies have 
neglected a possible nebular contribution which may be prominent at
cosmological redshifts.

\section{Emission measure of the recombination halo}
We assume the evolution of the background ionizing flux at the Lyman limit as
$J_{\rm UV}(z)=[(1+z)/3.5]^{\alpha} \times 10^{-21}$ erg cm$^{-2}$ s$^{-1}$
sr$^{-1}$ Hz$^{-1}$, and its frequency dependence at all redshifts as
(Chiba \& Nath 1997; Haardt \& Madau 1996)
$$J_{\rm UV} =J_{\rm UV}(z) \left( \frac{\nu}{\nu_0}  \right)^{-\beta}.\eqno (1)$$
This is the major force driving the redshift
evolution of emission measure $\epsilon_m$ of the ionized gas. We 
emphasize that this is just a working model, since precise normalization of
eq. (1) is still elusive, due to 1$\sigma$ uncertainties of factors of 6 or more
(Bajtlik, Duncan \& Ostriker 1988; Kulkarni \& Fall 1993;
Vogel et al. 1995). Further 
redshift dependence may come through chemical evolution which
influences cooling rate once $Z/Z_\odot > 0.01$ (B\"ohringer \& Hensler 1989).
This would imply slow $z$-evolution of the electron temperature, $T_e$. The 
dynamical evolution of the
disk--halo connection (through galactic fountain or some similar mechanism)
would also change intrinsic properties of the halo clouds. We neglect these
rather subtle points in this discussion. 

To proceed, we assume the fiducial $\log N_{\rm H\ I} = 17.4$ cm$^{-2}$ 
photoionized 
cloud residing in extended galactic halo of a galaxy at redshift $z$. 
This (in 
the first approximation, homogeneous) halo cloud of $\sim 30\, h^{-1}$ kpc
 in size is what we expect to see 
according to the metal-line absorption (Steidel 1993; Steidel et al.
1994; Petitjean \& Bergeron 1994). Galactic disks are regarded as 
opaque to the ionizing radiation between 1 and 4 Ryd. Our conclusions are 
valid as long as the cloud remains optically thin to H$\alpha$, which may
 not be true only at much higher column densities than those discussed here.

Emission measure of the fluorescent ${\rm Ly}\alpha$\ emission under the 
assumption of 
isothermal clouds at all epochs is given by
$$\epsilon_m ({\rm Ly}\alpha) 
= \frac{1.5 \times 10^2}{(2.75 + \beta)(1+z)^4}\left( 
\frac{N_{\rm H\ I}}{3\times 10^{17} \; {\rm cm}^{-2}} \right)\times$$
$$\, \, \, \, \, \; \; \;\; \; \left( \frac{1+z}{3.5} 
\right)^\alpha 
\left( \frac{T_e}{10^4 \; {\rm K}} \right)^{0.75} {\rm pc \; cm^{-6}}.\eqno (2)$$
where $T_e$ is
the electron temperature of the clouds (Osterbrock 1989; \'Cirkovi\'c \& 
Samurovi\'c 1998). 
This equation is valid for one-sided ionization of a hydrogen slab and is 
only valid for
$\log N_{\rm H\ I} \leq 17.4$ cm$^{-2}$. Once the slab thickness exceeds one 
optical depth 
for the ionizing photons, the emission measure depends only on the external
 ionizing flux.
If we denote the H$\alpha$/${\rm Ly}\alpha$\ ratio with 
$\delta$, we obtain the intensity in the H$\alpha$ line (Reynolds 1992):
$$
I_{{\rm H}\alpha}  =  1.44 \delta \left( \frac{T}{10^4 \; {\rm K}} 
\right)^{-0.92} \!
\epsilon_m ({\rm Ly}\alpha) \; {\rm R}$$  
$$\, \, \, \, \, \; \; \;\; \;  =  \frac{1.46}{(1+z)^2} \left( \frac{N_{\rm 
H\ I}}{3\times 10^{17} 
\; {\rm cm}^{-2}} \right) \left( \frac{T}{10^4 \;{\rm K}} \right)^{-0.17} \; 
{\rm R},\eqno(3)$$
($1 \; {\rm R}=10^6/4\pi$ photons cm$^{-2}$ s$^{-1}$ sr$^{-1}$) for a plausible
values of $\delta=0.08$, $\beta = 1.73$ (Haardt \& Madau 1996) and $\alpha
\simeq 2$ (Chiba \& Math 1997). This formula is valid for the equilibrium case
in which electronic and kinetic temperatures of the photoionized phase are
equal. Our result thus generalizes a  similar one obtained by
Bland-Hawthorn et al. (1994). The difference at a fiducial point
discussed by Bland-Hawthorn et al. (1994) can be attributed to
the different choice of metagalactic background spectral index and other 
not-well-known parameters.  Note that the largest uncertainty in the
equation (2) comes from the uncertainty in $\beta$, since both the models and 
observational results from the proximity effect exhibit a rather large scatter
(Bajtlik et al. 1988; Madau 1992; Kulkarni \& Fall 1993; Donahue
et al. 1995; Vogel et al. 1995; Haardt \& Madau 1996).

Several interesting approximate relationships can be obtained using eqs. (2)
and (3). The electron number density in fluorescing clouds is
roughly given by
$$n_e  =  \sqrt{\frac{\epsilon_m ({\rm H}\alpha)}{f l}}=$$ 
$$     =  \frac{0.0048
f}{1+z} \! \left( \frac{N_{\rm H\ I}}{3\times 10^{17}\; {\rm cm}^{-2}}
\right)^{\frac{1}{2}} \! \! \left( \frac{l}{10 \; {\rm kpc}} 
\right)^{-\frac{1}{2}} \! \!
\left( \frac{T_e}{10^4 \; {\rm K}} \right)^{0.375}\! \eqno(4)$$
where $n_e$ is given in ${\rm cm}^{-3}$ and  $f$ is the filling 
factor of the ionized gas (Reynolds 1987). For example, for a 
Lyman-limit system with $\log N_{\rm H\ I} =17.2$ cm$^{-2}$ and size of
 $l=30 \, h^{-1}$ kpc at low redshift, this formula (with, probably 
 unrealistic, assumption $f \approx 1$) gives $n_e \approx 5.7 
 \times 10^{-3}$ cm$^{-3}$ (using $h=0.75$ and $T_e \equiv T = 
 3\times 10^4$ K), both observationally allowed and theoretically 
 attractive value for highly photoionized regions giving rise to
  metal-line and Lyman-limit 
absorption systems. 

It will be of great interest to compare values obtained through equation 
(4) with those obtained by some independent procedure, say curve-of-growth
 measurements of abundance ratios of pairs of coupled metal species, or
  observations of Faraday screening of background radio-sources by a 
  foreground electron column density (Bland-Hawthorn
et al. 1995). 
Since the cosmological density parameter $\Omega$ and neutral hydrogen 
column densities at a given epoch are related (e.g.~\'Cirkovi\'c \& 
Samurovi\'c 1998), it should be possible
to establish what fraction of the cosmological density
is contained within the optically thin photoionized gas (eq. 3).

\section{Signal-to-noise as a function of redshift}

We now demonstrate the feasibility of detecting galaxy haloes in optical 
line emission at cosmological redshift.  We know from the Hubble Deep Field
that normal galaxies and galaxy haloes were in place by $z\sim 1$
(Steidel 1998).  Suppose that we
are observing target subtending solid angle $\Theta$ with the telescope of
diameter $D$, our detector is of efficiency $f$, and the total exposure 
time is $t$. Then, the total number of photons from the source in the 
H$\alpha$ line is
given as $n_{\rm source}=(\pi/4) D^2 \zeta f t \Theta I_{{\rm H} 
\alpha}$  
(Gould \& Weinberg 1996). The fraction of photons penetrating Earth's
 atmosphere is denoted by $\zeta$. The number of photons of the 
 background in the H$\alpha$ line profile in this case is approximately 
 equal to $n_{\rm sky} = (\pi/4)D^2 f
t \Theta \lambda/c (4 \sqrt{\pi} \sigma) \phi_{\rm sky}$ where $\lambda \equiv 
\lambda_0 (1+z)$ is the wavelength of the line centroid ($\lambda_0 = 6562.8$
\AA\  for the H$\alpha$ line), $ (4 \sqrt{\pi} \sigma)$ is the width of the 
line, 
determined, presumably, by thermal line-broadening, and $\phi_{\rm sky}$ is the
sky flux in photons cm$^{-2}$ s$^{-1}$ \AA$^{-1}$ arcsec$^{-2}$. Thus, 
signal-to-noise ratio is given by
$$S/N   =  \frac{\pi^{\frac{1}{4}}}{4} \zeta D \sqrt{\frac{f t
 \Theta}{\lambda_0  \sigma (1+z)}}\frac{I_{{\rm H}\alpha}}{\sqrt{\phi_{\rm
  sky}}} = $$

$$ = 1.7 \times 10^3 \zeta \left( \frac{f}{0.1} \right)^\frac{1}{2}
 \left( 
\frac{t}{4\; {\rm h}} \right)^\frac{1}{2} \! \left( \frac{D}{8\; {\rm m}} 
\right) \times$$ 
 $$  \left( \frac{\sigma}{20 \; {\rm km\; s}^{-1}} \right)^{-\frac{1}{2}} \! 
 \left( \frac{\Theta}{1+z} \right)^\frac{1}{2} \frac{I_{{\rm 
H}\alpha}}{\sqrt{\phi_{\rm sky}}}. \eqno(5)$$
We shall take the sky noise as constant and equal to the fiducial value at 
8250\AA\ (Offer \& Bland-Hawthorn 1998).  The redshift dependence of the 
actual observing wavelength changes
the background flux and it actually does increase
as we go at higher and higher redshifts (i.e. deeper in the infrared). 
We shall take the background flux to be 
$\phi_{\rm sky} = 5.44 \times 10^{-6}$ photons cm$^{-2}$ s$^{-1}$ \AA$^{-1}$
arcsec$^{-2}$. 
Our adopted value is sufficiently conservative as there are comparably
dark bands between the OH bandheads throughout the $R$, $I$, $z$, $J$
and $H$ bands. In our calculation, we neglect detector read noise: modern day
 optical 
and infrared detectors have amplifier noise of only 1$e^-$ and 5$e^-$ 
pix$^{-1}$
respectively, such that the sky background always dominates.

We note that for a long-slit of width $d$ (in arcsec) and
optimal placing, $\Theta \approx d \, \theta (l, z)$, where $\theta(l,z)$ is
 the angle
spanned by the target of characteristic proper size $l$ at redshift $z$. 
In general
case of FRW-universes with $\Lambda=0$, this angle is given by
(Weinberg 1972): 
$$\Theta ={lH_0\over{2c}} {(1+z)^2\Omega ^2\over {\Omega z}+ 
2 \left ( {\Omega\over 2} -1 \right ) \left ( \sqrt{1+\Omega z}-1
\right )} \equiv {lH_0\over{2c}} G(z), \eqno(6)$$
where we have denoted the redshift (and cosmological model) dependence by a
function $G(z)$. Assuming $\zeta \simeq 1$ and using the result obtained for
$I_{{\rm H}\alpha}$ in the equation (2), we may write 
$$ S/N  \approx  19 \left( \frac{f}{0.1} \right)^\frac{1}{2} \! \left( 
\frac{t}{4\; {\rm h}} \right)^\frac{1}{2} \! \left( \frac{D}{8\; {\rm m}} 
\right) \! \left( \frac{\sigma}{20 \; {\rm km\; s}^{-1}} 
\right)^{-\frac{1}{2}}\times$$ 
$$\! \left( \frac{N_{\rm H\ I}}{3\times 10^{17}\;
 {\rm cm}^{-2}} \right) \! 
  \! \left( 
\frac{T}{10^4\; {\rm K}} \right)^{-0.17} \! \left( \frac{l}{10\; {\rm kpc}} 
\right)^{\frac{1}{2}} 
\sqrt{\frac{G(z)}{(1+z)^5}}. \eqno(7)$$
\beginfigure{1}
\psfig{file=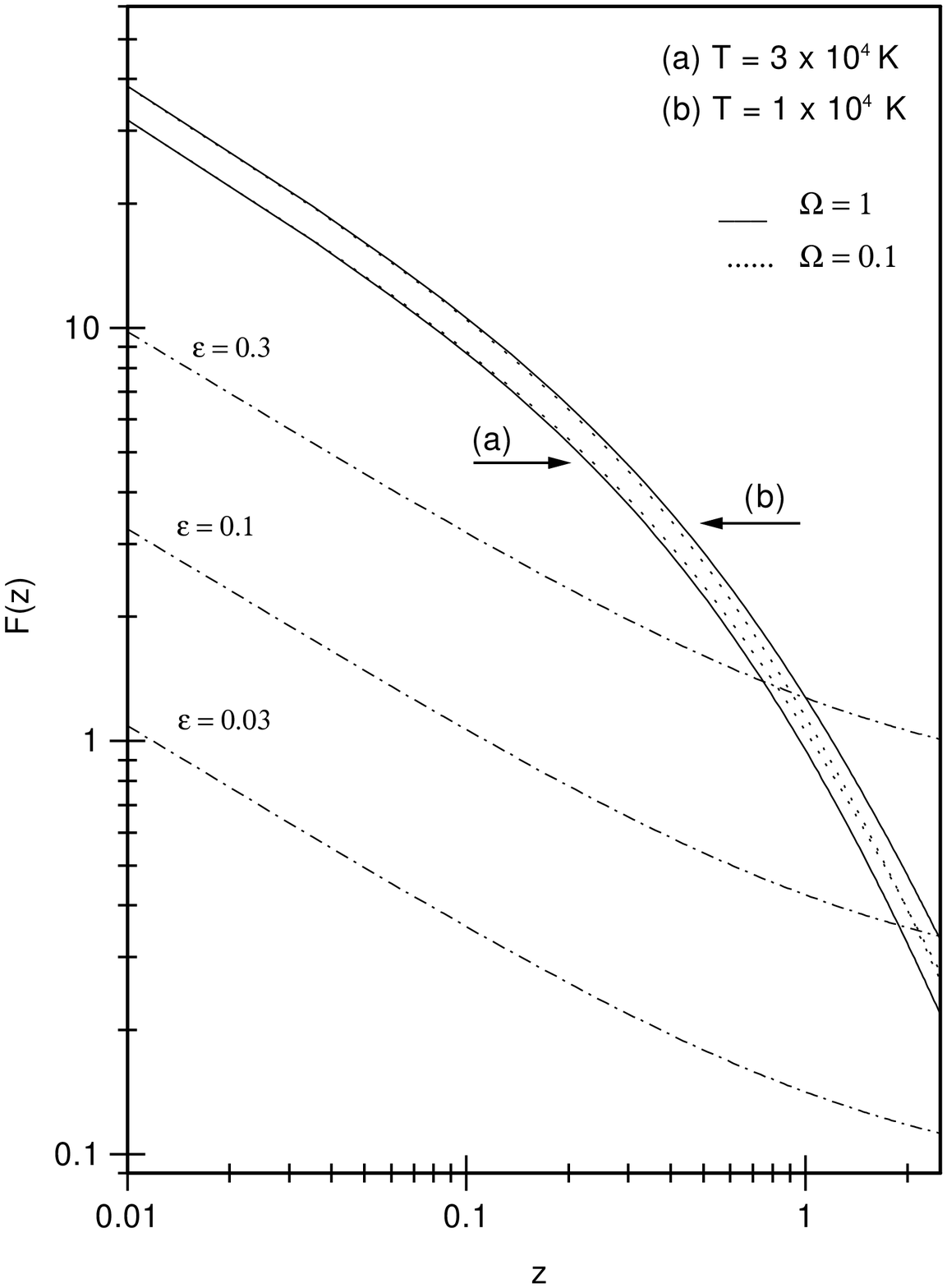,height=12cm}
\caption{{\bf Figure 1.}~The function $F(z) \equiv (S/N) \left( \frac{f}{0.1} 
\right)^{-1/2}
\left( \frac{t}{4\; {\rm h}} \right)^{-1/2}$ representing signal-to-noise
ratio predicted for a fiducial observing run with following parameters:
telescope diameter $D=8$ m, spherical target of the physical size $l=40\; 
h^{-1}$ kpc (which  is characteristic for metal-line absorbing haloes of 
$L_\ast$ galaxies); average column density (similar to the lower limit 
for optical thickness at the Lyman  
limit) $\log N_{\rm H\ I}=17.4$ cm$^{-2}$, velocity dispersion $\sigma 
= 20$ km s$^{-1}$  and two cases for the kinetic temperature $T=3 \times 
10^4$ K and $T=10^4$ K. Solid curve represents $F(z)$ for $\Omega=1$ and
 dotted for $\Omega=0.1$ universes (everywhere $h=0.75$).  
Dash-dotted lines represent (feasible) cases of fixed emission measures 
(for $T=3\times 10^4$ K) in $\Omega=1$ universe, given for comparison.} 

\endfigure
Note that this equation may be still simplified if we assume that 
thermal broadening dominates, such that $\sigma= \sqrt{2kT/\mu}$.
We consider two plausible  cosmologies $\Omega=1$ and $\Omega =0.1$
 (everywhere
$\Lambda =0$).

There are two basic approaches to obtaining very deep emission line
detections at optical and infrared wavelengths: long-slit spectroscopy and 
Fabry-Perot staring (e.g. Bland-Hawthorn et al. 1995).  {\it
It is essential to disperse the light over the field of view.}
The Fabry-Perot interferogram is dispersed radially over the detector
and reaches the deepest limits to date (1 mR at 1$\sigma$ in 6 hours for
a source that fills the field.) The power of the method comes, in part,
from the ability to disperse a small spectral window ($\sim 20-50$ \AA) 
over a large area detector.  The final spectrum is obtained from azimuthal
binning over the detector with typically $\sim 10^5$ pixels contributing
to each Angstrom bin. In comparison, the long-slit spectrum is dispersed 
along one axis of the detector.  Modern day spectrographs project the same
spectral window over orders of magnitude fewer pixels. The deepest diffuse
detections to date at the MSO 2.3m telescope with the Double Beam 
Spectrograph reach emission measures of 10 mR in
the same exposure time.\note * {\sevenrm At the same resolving power, the 
Gaussian profile of the slit in signal-to-noise gain
compared to the Fabry-Perot Airy profile.}

The shape of the emitting cloud is not important in the first
instance, as far as it does not change the optical depth for ionizing
radiation significantly. We plot the efficiency and exposure time-normalized
S/N as a function of redshift (Figure 1) for an observing run on 8m class
telescope for $l=40\, h^{-1}$ kpc halo cloud. The differences between 
cosmological models are negligible.  As far as lower redshift objects are
concerned, S/N $\simeq 10$ for a $\log N_{\rm H\ I} =16.5$ cm$^{-2}$ cloud
located at $z=0.5$ would, for instance, be achieved in a total exposure time of
$t \simeq 3$ h on a 8m telescope with efficiency $f=0.1$. 

In detecting diffuse
sources, accurate sky subtraction is critical since variable atmospheric (OH, 
water,
Fraunhofer lines) features can cause spurious positive signals. 
The best background subtraction is achieved when the sky is observed
simultaneously with the data, or almost simultaneously in the case of chopping.
In order to match 
the noise characteristics, there should be an equal number of sky and data pixels. 
The natural f/8 Cassegrain field for an 8m telescope is approximately 4 arcmin.
The redshift at which $\theta(l=40h^{-1}\; {\rm kpc},z)$ is equal to 4 arcmin is 
$z\approx 0.01$. At higher redshifts, galaxy haloes occupy only a fraction $x$ 
of the field of view.  Once the object fills more than half the
entrance aperture, say, $x$, you are better off rejecting $2(x-0.5)$ of
the data to match the sky solid angle, and there is no need to
chop until $x > 0.75$.  Thus, the transition between direct and interleaved
observation is actually a smooth one. 
At the Anglo-Australian Telescope, chopping is achieved successfully for any
spectrograph with charge shuffling synchronized to the nodding of the telescope.

\section{Discussion}
Halo gas in galaxy haloes at cosmological distances has traditionally been
studied using resonance absorption lines in the spectra of background quasars.
However, we find that, in certain instances, H$\alpha$ detections may be possible 
for metal-line absorbing gas around luminous galaxies out to $z\sim 1$.
Our results are contingent upon the assumption that the major force
governing the evolution of extended gas around galaxies is the evolving
metagalactic ionizing background. Generalized formulae for the emission measure
of recombining uniform haloes are given. It is shown that surface brightness of such
objects does not conform to a simple $S \propto (1+z)^{-4}$ relationship. One 
can, 
in principle, measure the deviation of H$\alpha$ intensity from eq. (3) and thus 
trace the intrinsic halo evolution, but only if one can account
successfully for transition of baryons from the diffuse gaseous phase to other
forms of baryonic matter (stars, molecular clouds, etc.). Since only the
integrated neutral column density along the line of sight through a halo comes
into play, we agree with Hogan and Weymann (1987) that the feasibility is
essentially independent of the exact model of halo clouds. 

It is important to add that results
for intensity and signal-to-noise ratio of the fluorescing gas at any given
redshift should be regarded as the lower limits, for at least two reasons. 
Apart from the background ionization, there are probably other ionizing sources
(Donahue et al.  1995; Fukugita et al. 1998), like leakage of small 
fraction of Lyman-limit photons from
galactic disks, which is small at present  (Leitherer et al. 1995;
Bland-Hawthorn  1997), but could have been much higher in the past (Giroux \&
Shull 1997), depending on the timescales for dust formation; other possible
influences are ionizing intergalactic shocks, halo (or
intergalactic) stars, or even the decaying dark matter
(Sciama 1993, 1995). Secondly, the ratio $\delta$ seems to be consistently higher in
real systems than in Case B recombination calculations assumed here (see
Charlot \& Fall 1993; Bechtold et al. 1997 and references therein). A
realistic halo model should be able to quantify these effects 
and determine exact ionization structure of the extended galactic gas. Our
results include no correction for reddening (e.g. Tresse \& Maddox 1997), since we 
assume dust--free gaseous haloes and observing far from the Galactic plane.  

Such experiments should be able to answer a basic question: to what extent
is the recombining gaseous halo a generic phenomena in galaxies?  A successful
detection of diffuse H$\alpha$ emission from haloes of normal galaxies at low  
redshifts will have enormous theoretical and practical significance. It will
enable direct testing of models for the evolution of the metagalactic ionizing
flux. It will also determine what fraction of halo light at intermediate and 
low 
redshift is continuum and what fraction is contained in the emission lines
(Charlot \& Silk 1995). Finally, if halo emission is as luminous as we 
anticipate,
a broad spectral campaign could constrain physical conditions in extended
galactic haloes (or disks), e.g. proton density (Reynolds 1987), pressure 
(Bowyer et al. 1995), clumpiness,  and so on.


\section*{References}
\beginrefs

\bibitem Adams, F.C.,  Lauglin, G., 1996, ApJ, 468, 586 

\bibitem Bajtlik, S. Duncan, R. C.,  Ostriker, J. P., 1988, ApJ, 327, 570

\bibitem Bechtold, J., Yee, H. K. C., Elston, R.,  Ellingson, E., 1997,
{ApJ}, {477}, L29

\bibitem Binette, L., Wang, J. C. L., Zuo, L.,  Magris, C. G., 1993,
{AJ}, {105}, 797

\bibitem Bland-Hawthorn, J., 1997, {PASA}, {14}, 64

\bibitem Bland-Hawthorn, J., Taylor, K., Veilleux, S.,  Shopbell,
P. L., 1994, {ApJ}, {437}, L95

\bibitem Bland-Hawthorn, J., Ekers, R. D., van Bruegel, W., Koekemoer,
A.,  Taylor, K., 1995, {ApJ}, {442}, L77

\bibitem Bland-Hawthorn, J., Freeman, K. C.,  Quinn, P. J., 1997, {ApJ},
{490}, 143

\bibitem B\"ohringer, H.,   Hensler, G., 1989, {A\&A}, {215}, 147

\bibitem Bowyer S., Lieu, R., Sidher, S. D., Lampton, M.,  Knude, J., 1995, 
{Nat}, {375}, 212

\bibitem Bregman, J. N., 1981, {ApJ}, {250}, 7

\bibitem Brunner, R. J., Connolly, A. J., Szalay, A. S.,  Bershady,
M. A., 1997, ApJ, 482, L21

\bibitem Carr, B., 1994, {ARA\& A}, {32}, 531

\bibitem Chabrier, G.,  Mera, D., 1997, A\& A, 328, 83

\bibitem Charlot, S.,  Fall, S. M., 1993, {ApJ}, {415}, 580

\bibitem Charlot, S.,  Silk, J., 1995, {ApJ}, {445}, 124

\bibitem Chen, H.-W., Lanzetta, K. M., Webb, J. K.,  Barcons, X., 1998, 
{ApJ}, 498, 77

\bibitem Chiba, M.,  Nath, B. B., 1997, {ApJ}, {483}, 638

\bibitem Connolly, A. J., Csabai, I., Szalay, A. S., Koo, D. C., Kron, R. G.,
 Munn, J. A., 1995, AJ, 110, 2655

\bibitem \'Cirkovi\'c, M. M.,  Samurovi\'c, S., 1998, {Ap\&SS}, 257, 95

\bibitem Donahue, M., Aldering G.,   Stocke, J. T., 1995, {ApJ}, {450}, L45

\bibitem Flynn, C., Gould, A.,  Bahcall, J. N., 1996, {ApJ}, {466}, L55

\bibitem Fukugita, M., Hogan, C. J.,  Peebles, P. J. E., 1998, 
{ApJ},  503, 518

\bibitem
Giroux, M. L. and Shull, J. M., 1997, {AJ}, {113}, 1505

\bibitem Glazebrook, K., Offer, A. R.,  Deeley, K., 1998, {ApJ}, {492}, 98

\bibitem Gould, A.,  Weinberg, D. H., 1996, {ApJ}, {468}, 462

\bibitem Haardt, F.,  Madau, P., 1996, {ApJ}, {461}, 20

\bibitem Hogan, C. J.,  Weymann, R. J., 1987, {MNRAS}, {225}, 1{\sixrm P}

\bibitem Kovalenko, I. G., Shchekinov, Iu. A.,  Suchkov, A. A., 1989, 
{Ap\&SS}, {152}, 223

\bibitem Kulkarni, V. P.,  Fall, S. M., 1993, ApJ, 413, L63

\bibitem Lanzetta, K. M., Yahil,  A.,  Fern\'andez-Soto, A., 1996, 
{Nat}, {381}, 759

\bibitem Lanzetta, K. M., Bowen, D. V., Tytler, D.,   Webb, J. K., 1995,
{ApJ}, {442}, 538 

\bibitem Leitherer, C., Ferguson, H. C., Heckman, T. M.,  Lowenthal,
J. D., 1995, ApJ, 454, L19

\bibitem Madau, P., 1992, ApJ, 389, L1

\bibitem Maloney, P., 1992, {ApJ}, {398}, L89

\bibitem Mo, H. J., 1994, {MNRAS}, {269}, L49

\bibitem Mo, H. J.,  Miralda-Escud\'{e}, J., 1996, {ApJ}, {469}, 589

\bibitem Moore, B., 1993, ApJ, 413, L93

\bibitem Navarro, J. F.,  White, S. D. M., 1994, {MNRAS}, {267}, 401

\bibitem Norman, C. A.,   Ikeuchi, A., 1989, {ApJ}, {345}, 372

\bibitem Nulsen, P. E. J.,  Fabian, A. C., 1997, MNRAS, 291, 425 

\bibitem Offer, A. R.,  Bland-Hawthorn, J., 1998, MNRAS, 299, 176

\bibitem Osterbrock, D. E., 1989, {Astrophysics of Gaseous Nebulae and
Active Galactic Nuclei}, University Science Books, Mill Valley

\bibitem Petitjean, P.,  Bergeron, J., 1994, A\&A, 283, 759

\bibitem Reynolds, R. J., 1992, {ApJ}, {392}, L35 

\bibitem Reynolds, R. J., 1987, {ApJ}, {323}, 553

\bibitem Richstone, D., Gould A., Guhathakurta,  Flynn, C., 1992,
{ApJ}, {388}, 354 

\bibitem Sawicki, M. J., Lin, H.,  Yee, K. C., 1997, {AJ}, {113}, 1

\bibitem Sciama, D.  W., 1993, {Modern Cosmology and 
the Dark Matter Problem}, Cambridge Univ. Press, Cambridge

\bibitem Sciama, D. W., 1995, {MNRAS}, {276}, L1

\bibitem Shectman, S. A., Landy, S. D., Oemler, A., Tucker, D. L., Lin, H.,
Kirshner, R. P.,  Schechter, P. L., 1996, {ApJ}, {470}, 172

\bibitem Steidel, C.  C., 1993, in Majewski, S. R., ed., 
{Galaxy Evolution: The Milky Way
Perspective\/},  ASP Conf. Ser. Vol. 49, 
Astron. Soc. Pac., San Francisco,
p. 227

\bibitem Steidel, C. C., 1998, in Zaritsky, D., ed., 
{Galactic Halos: a UC 
Santa Cruz workshop}, ASP Conf. Ser. Vol. 136, 
Astron. Soc. Pac., San Francisco, p. 167

 

\bibitem Steidel, C. C., Dickinson, M.,  Persson, S. E., 1994, {ApJ}, {437}, L75

\bibitem Tresse, L.,  Maddox, S. J., 1998, {ApJ}, 495, 691

\bibitem Vogel, S. N., Weymann, R., Rauch, M.,  Hamilton, T., 1995, ApJ, 441, 
162

\bibitem Weinberg, S., 1972, {Gravitation and Cosmology}, 
 John Wiley  and Sons, New York, NY

\bibitem Wolfire, M. G., McKee, C. F., Hollenbach, D.,  Tielens, A. G. G. 
M., 1995, {ApJ}, {453}, 673

\bibitem Yahata, N.,  Lanzetta, K. M., 1999, in preparation
\bye